\documentclass[prl,preprint,groupedaddress,showpacs]{revtex4}

\usepackage{amsmath}
\usepackage{amssymb}
\usepackage{amsfonts}
\usepackage{epsfig}
\usepackage{bm}
\begin{document}

\title{Kinetics of the collisionless expansion of spherical nanoplasmas}

\author{F. Peano$^{1}$}
\author{F. Peinetti$^{1}$}
\author{R. Mulas$^{1}$}
\author{G. Coppa$^{1}$}\email{gianni.coppa@polito.it}
\author{L. O. Silva$^{2}$} \email{luis.silva@ist.utl.pt}
\affiliation{$^1$Dipartimento di Energetica, Politecnico di Torino, 10129 Torino, Italy}\author{ }
\affiliation{$^2$GoLP/Centro de F\'isica dos Plasmas, Instituto Superior T\'ecnico, 1049-001 Lisboa, Portugal}

\date{\today}

\begin{abstract}
The collisionless expansion of spherical plasmas composed of cold ions and hot electrons is analyzed using a novel kinetic model, with special emphasis on the influence of the electron dynamics. Simple, general laws are found, relating the relevant expansion features to the initial conditions of the plasma, determined from a single dimensionless parameter. A transition is identified in the behavior of the ion energy spectrum, which is monotonic only for high electron temperatures, otherwise exhibiting a local peak far from the cutoff energy.
\end{abstract}

\maketitle

\newcommand{\ped}[1]{_{\text{#1}}}
\newcommand{\api}[1]{^{\text{#1}}}
\newcommand{\diff}[1]{\text{d}#1}

Recent experiments on the interaction of ultraintense laser pulses with atomic and molecular clusters \cite{cluster_exp,Sakabe} have shown the possibility of accelerating ions  to energies of interest for many applications, such as nuclear fusion and X-ray generation. The interpretation of these experiments and the need of controlling critical features of the ion acceleration (in particular, their energy spectrum) require a theoretical insight of the laws governing the collisionless expansion of a spherical plasma driven by hot electrons. A detailed knowledge of the kinetics of the expansion is also necessary for particular applications, such as the biomolecular imaging with ultrashort X-ray pulses \cite{xrays}, where sample damage before the imaging time must be avoided. Moreover, the strong interplay between ions and electrons when using high-intensity lasers can be used to control the expansion of large clusters, by tailoring the ion phase space \cite{Peano}.
At present, analytical solutions for a spherical expansion exist only for ideal cases, such as the Coulomb explosion (CE) \cite{Kaplan_PRL} of a pure ion plasma, which occurs when all the electrons are suddenly swept away from the cluster by the laser field. In opposite conditions, hydrodynamic models have been proposed \cite{hydrodynamic} to estimate basic features of expansions far from the CE regime, while a kinetic solution for the adiabatic expansion of plasma bunches into a vacuum has been derived in the quasineutral limit \cite{Kovalev}. However, such extreme situations are hardly met in experiments, in particular for large clusters, as the laser intensity is insufficient to drive a pure CE but high enough to produce relevant charge buildups within the clusters: in such cases, the expansion is described correctly only by kinetic models, based on the Vlasov-Poisson equations. In this Letter, the long-term dynamics of the expansion of spherical plasmas is analyzed by using a novel Lagrangian model, which allows a self-consistent, kinetic description of the radial motion of the ions and of the three-dimensional motion of nonrelativistic electrons, with high accuracy and low computational effort. The important case of initially-Maxwellian electrons is investigated in detail, with special emphasis on the influence of the initial conditions. Simple relationships are deduced for the most important physical quantities, valid for a wide range of characteristic parameters of the expansion. These empirical formulae are surprisingly accurate and suggest the possibility of a rigorous analytical derivation. Furthermore, the study of the asymptotic phase of the expansion shows that the behavior of the ion energy spectrum for realistic conditions differs greatly from that of a pure CE. Finally, the importance of these results to interpret experimental data is discussed.

For the present analysis, the electrons are assumed to be heated instantaneously, as upon irradiation with an ultrashort laser pulse, and their initial distribution is assumed to be Maxwellian, with temperature $T_0$. The expansion process is divided in two stages: a rapid expansion of the electrons, which leads to an equilibrium configuration before the ions move appreciably, and a subsequent, slower expansion of the plasma bulk. The analysis of the early expansion (characterized by density oscillations that are damped on a time scale much faster than the ion motion) is necessary in order to evaluate the equilibrium distribution of the electrons, which is used as initial condition for the bulk expansion.
The dynamics of the plasma is analyzed self-consistently by following the motion of the ions, along with the evolution of the electron distribution, which is described as a sequence of ergodic equilibrium configurations, so that the only independent variable related to the electrons is their total energy $\epsilon=\frac{1}{2}m\ped{e}\mathbf{v}^2-e\Phi$ (the electrostatic potential $\Phi$ is set to zero at infinity, so that $\epsilon<0$ for trapped electrons). A Lagrangian approach is used for both ions (which move radially, starting from the initial position $r_0$, with zero velocity) and electrons (whose energy $\epsilon$ evolves in time starting from the initial value $\epsilon_0$). The model determines the ion trajectories $r\ped{i}(r_0,t)$, the electron energies $\epsilon(\epsilon_0,t)$, the ion density $n\ped{i}(r,t)$, the electron density $n\ped{e}(r,t)$, and the potential $\Phi(r,t)$, according to the set of equations
\begin{align}
	&m\ped{i} \dfrac{\partial^2 r\ped{i}}{\partial t^2} = -Ze \dfrac{\partial \Phi}{\partial r}(r\ped{i})\tag{1a}\label{eq:model_a}\\
	&\dfrac{1}{r^2}\dfrac{\partial}{\partial r}\left(r^2\dfrac{\partial \Phi}{\partial r}\right) = 4\pi e\left(n\ped{e}-Zn\ped{i}\right)\tag{1b}\label{eq:model_b}\\
	&n\ped{i}(r\ped{i}) = n\ped{i,0}(r_0)\dfrac{r_0^2}{r\ped{i}^2}\!\Big/\dfrac{\partial r\ped{i}}{\partial r_0}\tag{1c}\label{eq:model_c}\\
	&n\ped{e} = \displaystyle\int\! \rho\ped{e,0}(\epsilon_0) \frac{\sqrt{\epsilon+e\Phi}}
{4\pi\int\! {r^\prime}^2\sqrt{\epsilon+e\Phi(r^\prime)}\diff r^\prime} \diff \epsilon_0 \tag{1d}\label{eq:model_d}\\
	& \frac{d}{dt} \left[\frac{32\sqrt{2}}{3}\pi^2m\ped{e}^{3/2} \int \left(\epsilon+e\Phi\right)^{3/2} r^2 \diff r \right] = 0
\tag{1e}\label{eq:model_e}
\end{align}
where $m\ped{i}$ is the ion mass, and $Z$ the ion charge state. The expansion dynamics is determined once the initial ion density $n\ped{i,0}$ and the electron energy distribution $\rho\ped{e,0}$ are given.
In Eq. \eqref{eq:model_d}, the electron density is expressed as the sum of the number of electrons $\rho\ped{e}(\epsilon)\diff\epsilon = \rho\ped{e,0}(\epsilon_0)\diff\epsilon_0$ having energy in $[\epsilon,\epsilon+\diff\epsilon]$, multiplied by the probability for an electron with energy $\epsilon$ to be found at the radius $r$, according to the ergodic distribution.
Due to the large mass disparity between ions and electrons, the ergodic invariant \cite{Ott} for the electrons [i.e., the phase-space volume enclosed by the surface of equation $\frac{1}{2}m\ped{e}\mathbf{v}^2-e\Phi = \epsilon$, as defined in Eq. \eqref{eq:model_e}] is conserved; this determines the electron energy distribution at any time during the expansion of the plasma bulk.
Even though Eq. \eqref{eq:model_c} is written under the hypothesis of no ion overtaking ($\partial r\ped{i}/\partial r_0 \neq 0$) \cite{Kaplan_PRL}, the model can be easily generalized to include many-branched shock shells \cite{Peano} and different ion species.
As can be noticed by writing Eqs. (1) in dimensionless form, the dynamics of the expansion depends on the single parameter $\hat{T}_0 = Zk\ped{B}T_0/\epsilon\ped{CE}=3\lambda\ped{D}^2/R_0^2$, being $\lambda\ped{D}$ the Debye length for the electrons, and $\epsilon\ped{CE}=ZeQ_0/R_0$ the maximum ion energy attainable from the CE of a uniformly-charged sphere of ions, with radius $R_0$ and total charge $Q_0$.

\begin{figure}[!b]
\centering \epsfig{file=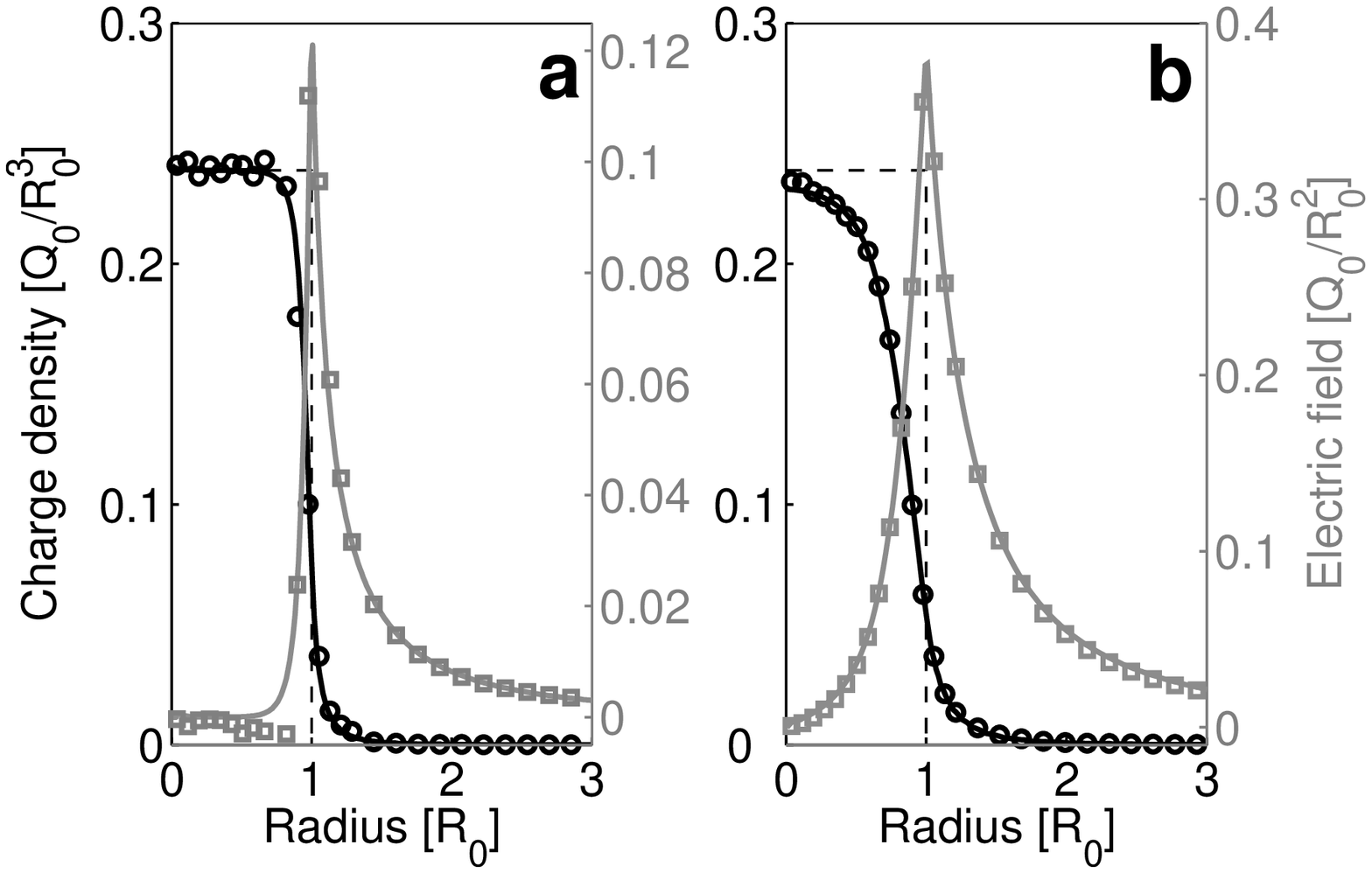, height=2in}
\caption{Electron density (black) and electric field (gray) for (a) $\hat{T}_0= 7.2\times 10^{-3}$ and (b) $\hat{T}_0= 7.2\times 10^{-2}$, after the early electron expansion with immobile ions. Lines refer to the present theory and markers to particle-in-cell simulations.}
\label{nE}
\end{figure}

\begin{figure}[!htb]
\centering \epsfig{file=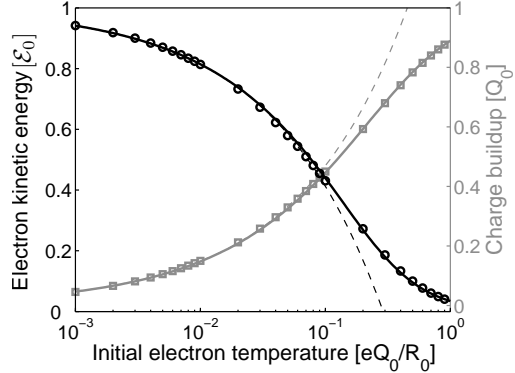, height=2in}
\caption{Charge buildup $\Delta Q$ (gray) and total electron kinetic energy $\mathcal{E}$ (black) as functions of $\hat{T}_0$, after the early electron expansion with immobile ions. Circles refer to the present theory, while solid lines refer to the fit laws of Eqs. \eqref{eq:fit1} and \eqref{eq:fit1b}. Dashed lines show the power-law behavior of $\Delta Q$ and $\mathcal{E}$ in the low-temperature regime.}
\label{qftf}
\end{figure}

The study of the bulk expansion does not require the detailed knowledge of the early expansion transient, but only its final equilibrium. Equations \eqref{eq:model_d} and \eqref{eq:model_e} are valid only for sufficiently-smooth variations of $\Phi(r,t)$ in time, a condition which is not met in the early stage when the electrons are suddenly allowed to expand (as if a rigid wall, initially confining the hot electrons, were instantaneously brought to infinity). However, these equations can still be used by introducing a rigid potential barrier and gradually moving it outward, provided that special care is taken to avoid any energy exchange between the electrons and the expanding wall [this is done with a suitable adjustment of Eq. \eqref{eq:model_e}], which would lead to an overestimate of the electron cooling.
The validity of this procedure has been tested by using reference results from particle-in-cell simulations \cite{codes}. In the comparisons, a cluster of radius $R_0=32$ nm was considered, with electron density $2.3\times 10^{22}$ cm$^{-3}$, for different electron temperatures spanning the range 1-10 keV.
In Figure \ref{nE}, the equilibrium electron density and the electric field are compared for $T_0 = 1$ keV ($\hat{T}_0= 7.2\times 10^{-3}$) and $10$ keV ($\hat{T}_0= 7.2\times 10^{-2}$), showing the excellent agreement between the different calculations. During the fixed-ion expansion, 5\% of the electrons leave the ion core for $\hat{T}_0= 7.2\times 10^{-3}$, while the percentage rises to 38\% when $\hat{T}_0= 7.2\times 10^{-2}$.
Figure \ref{qftf} shows the equilibrium charge buildup within the ion core, $\Delta Q$, and the total kinetic energy of the electrons, $\mathcal{E}$ (accounting for all confined electrons, inside and outside the ion core), as a function of $\hat{T}_0$.
Considering that the analytical results \cite{Crow} for semi-infinite planar expansions must be recovered in the spherical case for $\hat{T_0}\ll 1$ (as $\lambda\ped{D}\ll R_0$), a simple fit for $\Delta Q$ has been found, in the form
\begin{equation}
\Delta Q/Q_0 =  \mathcal{F}_{2.60}\left(\sqrt{6/e} \ \hat{T}_0^{1/2}\right)\text{,}
\label{eq:fit1}
\end{equation}
where $\mathcal{F}_\mu(x) = x/\left(1+x^\mu\right)^{1/\mu}$. As shown in Fig. \ref{qftf}, the accuracy of the fit is excellent even when more than 90\% of the electrons leave the ion core.
A similar fit holds for $\mathcal{E}$, as 
\begin{equation}
\mathcal{E}/\mathcal{E}_0 = 1-\mathcal{F}_{3.35}\left(1.86\:\hat{T}_0^{1/2}\right)\text{,}
\label{eq:fit1b}
\end{equation}
where $\mathcal{E}_0$ is the initial thermal energy of the electrons. In the low-temperature limit ($\hat{T_0}\ll 1$), Eqs. \eqref{eq:fit1} and \eqref{eq:fit1b} reduce to $\Delta Q/Q_0 \simeq \sqrt{6/e} \hat{T_0}^{1/2}$ (as in the semi-infinite planar expansion \cite{Crow}) and $\mathcal{E}/\mathcal{E}_0 \simeq 1 - 1.86\:\hat{T}_0^{1/2}$, respectively.

\begin{figure}[!htb]
\centering \epsfig{file=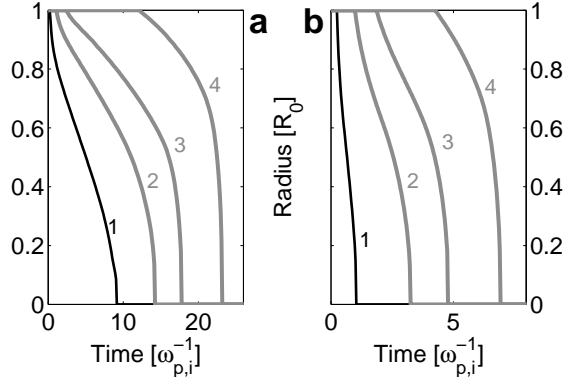, height=2in}
\caption{Radial trajectories of the isodensity points corresponding to 98\%, 75\%, 50\%, and 25\% of the initial value (curves 1 through 4), for (a) $\hat{T}_0= 7.2\times 10^{-3}$ and (b) $\hat{T}_0= 7.2\times 10^{-2}$. Times are in units of the inverse of the ion plasma frequency $\omega\ped{p,i}=\left(4\pi n\ped{i,0}Z^2e^2/m\ped{i}\right)^{1/2}$.}
\label{R_el}
\end{figure}

\begin{figure}[!htb]
\centering \epsfig{file=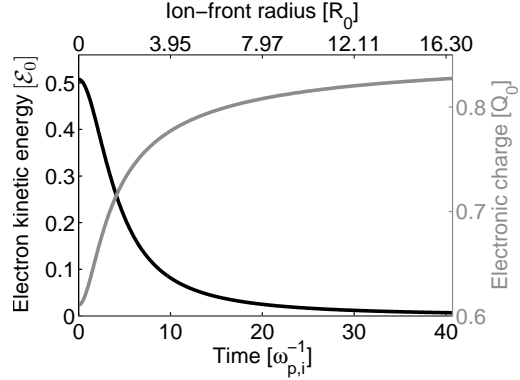, height=2in}
\caption{Evolution of the electronic charge within the ion front (gray), and of the total kinetic energy of the electrons (black), for $\hat{T}_0 = 7.2\times10^{-2}$.}
\label{qt}
\end{figure}

\begin{figure}[!htb]
\centering \epsfig{file=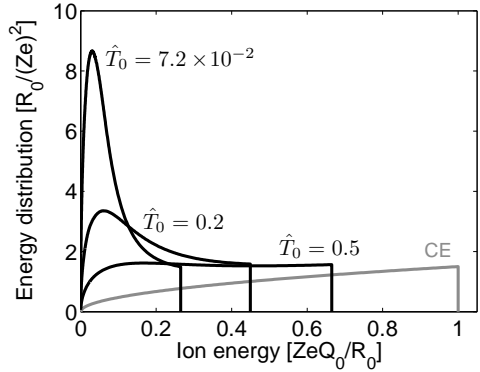, height=2in}
\caption{Asymptotic ion energy spectra (black) for $\hat{T}_{0}=7.2\times 10^{-2}$, $0.2$, and $0.5$. The gray curve refers to the pure CE case.}
\label{ionspectrum2}
\end{figure}

\begin{figure}[!htb]
\centering \epsfig{file=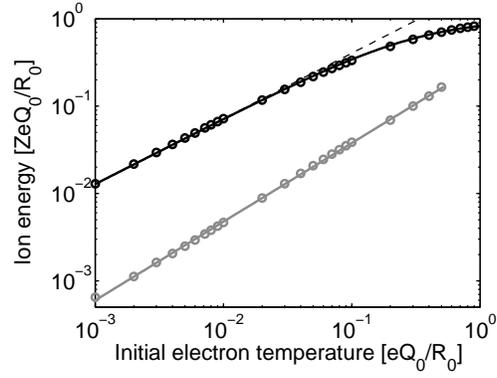, height=2in}
\caption{Cutoff ion energy (black) and position of the maximum in the ion energy spectrum (gray) as a function of $\hat{T}_0$: circles refer to the present theory, while the solid lines refer to the fit laws in the text. The dashed line represents the power-law behavior of $\epsilon\ped{max}$ for $\hat{T}_0 \ll 1$.}
\label{energy}
\end{figure}

Once the initial equilibrium distribution of electrons is determined, thus setting the initial conditions of the plasma, the expansion can be analyzed by solving Eqs. (1). The early phase of the bulk expansion is responsible for the main part of the ion acceleration. In this stage, the inner ions expand much slower than the outer ions: for $\hat{T}\ped{0} \ll 0.1$, the plasma core stays initially still, while a rarefaction front propagates inward until it finally encompasses the whole ion distribution. On the contrary, for $\hat{T}\ped{0} \gtrsim 0.1$, all the ions are promptly involved in the expansion. The different behaviors are depicted in Fig. \ref{R_el}, where the radial trajectories of isodensity points are plotted, for the same cases of Fig. \ref{nE}.
The case of Fig. \ref{R_el}b represents a reference situation of practical interest, in which the clusters undergo a rapid expansion in a hybrid regime, far from both the hydrodynamic limit and the CE regime. In such situation, previous models do not provide an accurate description of the expansion, because a self-consistent study of the electron and ion dynamics is required. As illustrated in Fig. \ref{qt} (for the scenario of Fig. \ref{R_el}b),
the electrons rapidly cool down and the charge buildup within the ion front decreases, until a ballistic regime is reached for both species \cite{Manfredi}.
In general, the electron dynamics strongly affects the energy spectrum of the ions, by reducing the repulsive electric field. Consequently, the final energies of the ions are lower with respect to the CE case. In fact, the asymptotic energy $\epsilon_\infty$ of an ion starting at $r_0$ is given by
\begin{equation}
\frac{\epsilon_\infty(r_0)}{Ze} = \frac{q(r_0,0)}{r_0} + \int^{\infty}_{0} \! \!  \frac{1}{r\ped{i}(r_0,t)}\frac{\partial q\left(r\ped{i}(r_0,t),t\right)}{\partial t}\diff{t}
\label{eq:E_ion} \text{,}
\end{equation} 
where $q(r,t)$ is the net charge buildup enveloped by a sphere of radius $r$ at time $t$. The integral term (vanishing for a CE) accounts for the energy loss due to the decrease of the positive charge buildup experienced by each ion along its trajectory. 
As expected, the shape of the asymptotic energy spectrum of the ions and its cutoff energy depend strongly on the initial conditions, as shown in Fig. \ref{ionspectrum2}: for $\hat{T}_0=7.2\times10^{-2}$, the spectrum exhibits a maximum at 12\% of the energy cutoff, $0.27\epsilon\ped{CE}$. Such distribution is qualitatively and quantitatively different from the asymptotic spectrum of a pure CE, $\frac{3}{2}(\epsilon/\epsilon\ped{CE})^{1/2}$, with cutoff energy $\epsilon\ped{CE}$. As $\hat{T}_0$ increases, the distribution flattens until the maximum disappears for $\hat{T}_0 = 0.5$ (a condition close to the CE regime, with cutoff above $0.7\epsilon\ped{CE}$), which can hence be taken as a lower bound for the validity of the CE model.
The cutoff ion energy $\epsilon\ped{max}$ as a function of $\hat{T}_0$ is shown in Fig. \ref{energy}, along with the energy corresponding to the peak in the spectrum, $\epsilon\ped{peak}$. The cutoff energy admits a simple fit in the form 
\begin{equation}
\epsilon\ped{max}=\mathcal{F}_{1.43}\left(2.28\:\hat{T}_0^{3/4}\right)\epsilon\ped{CE}\text{,}
\label{eq:fit2}
\end{equation}
which, for $\hat{T_0}\ll 1$, reduces to $\epsilon\ped{max}\simeq 2.28\:\hat{T}_0^{3/4}\epsilon\ped{CE}$, while, for $\hat{T}_0 < 0.5$, $\epsilon\ped{peak}$ exhibits a simple power-law behavior, as $\epsilon\ped{peak}=0.3\hat{T}_0^{0.9}\epsilon\ped{CE}$.

In recent experiments \cite{Sakabe} on the interaction of intense lasers ($10^{16}-10^{17}$ W/cm$^2$) with large hydrogen clusters (up to $2 \times 10^{5}$ atoms), the measured ion spectra exhibit a local maximum, which has been explained, in the framework of a CE model, as an effect of the distribution of cluster sizes.
The present analysis shows that nonmonotonic energy spectra can also arise from single-cluster effects, indicating that a maximum can appear even for narrow distributions of cluster radii (for example, with large clusters containing $10^6-10^7$ atoms). In such cases, the spectrum features (specifically, the laws for $\epsilon\ped{peak}$ and $\epsilon\ped{max}$ presented here) also provide an estimate for the initial electron temperature. 

In conclusion, the dynamics of the electron-driven expansion of spherical nanoplasmas has been fully analyzed using a new kinetic model. Simple, accurate laws have been derived for the general properties of the expansion, which are valid for any value of initial electron temperature (as long as relativistic effects are negligible). 
The study also revealed peculiar features of the ion energy spectrum (namely, the presence of a local maximum and the transition from nonmonotonic to monotonic behavior) that are not taken into account by simplified models and that can be important for the interpretation of experiments, where single-cluster effects are relevant. Moreover, the strong dependence on the initial conditions indicates the possibility of tailoring the ion spectrum, thus improving control over the plasma expansion.

\begin{acknowledgments}
Work partially supported by ASP (Italy) and by FCT (Portugal). The authors would like to acknowledge Prof. Ricardo Fonseca and Michael Marti for help with the OSIRIS simulations, performed at the expp cluster at IST, Lisbon.
\end{acknowledgments}

\end{document}